\documentclass[times,twocolumn,final]{elsarticle}
\usepackage{medima}
\usepackage{framed,multirow}

\usepackage{amssymb}
\usepackage{latexsym}

\usepackage{url}
\usepackage{xcolor}
\usepackage[hidelinks]{hyperref}
\definecolor{newcolor}{rgb}{.8,.349,.1}

\usepackage{mathtools}
\usepackage{amsmath}

\newcommand\Tstrut{\rule{0pt}{2.0ex}}         % = `top' strut
\newcommand\Bstrut{\rule[-0.9ex]{0pt}{0pt}}   % = `bottom' strut

\usepackage[ruled,vlined,algo2e]{algorithm2e}
\SetKwComment{Comment}{$\triangleright$\ }{}

\usepackage{hyperref}

\journal{Medical Image Analysis}

\begin{document}

\verso{B. Zhou \textit{et~al.}}

\begin{frontmatter}

%\title{Structural-guided Multi-Modal Registration by Learning Segmentation without Ground Truth: Application to Intraprocedural CBCT/MR Liver Segmentation and Registration}
% \title{Deep Feature Transformation Network for Personalized Federated Learning in Multi-institutional low-count PET Denoising}

\title{FedFTN: Personalized Federated Learning with Deep Feature Transformation Network for Multi-institutional Low-count PET Denoising}

% \author[1]{Bo \snm{Zhou}\corref{cor2}}
% \ead{bo.zhou@yale.edu}
% \author[1]{Huidong \snm{Xie}}
% \author[1]{Qiong \snm{Liu}}
% \author[1]{Xiongchao \snm{Chen}}
% \author[1]{Xueqi \snm{Guo}}
% \author[4]{Zhicheng \snm{Feng}}
% \author[5]{S. Kevin \snm{Zhou}}
% \author[1,2,3]{James S. \snm{Duncan}}
% \author[1,2]{Chi \snm{Liu}\corref{cor2}}
% \corref{cor1}
% \cortext[cor1]{Parts of the data used in the preparation of this article were obtained from the University of Bern, Department of Nuclear Medicine and School of Medicine, Ruijin Hospital. As such, the investigators contributed to the design and implementation of DATA and/or provided data but did not participate in the analysis or writing of this report. A complete listing of investigators can be found at: “https://ultra-low-count-pet.grand-challenge.org/Description/”}
% \cortext[cor2]{Corresponding author.}
% \ead{chi.liu@yale.edu}

\author[1]{Bo \snm{Zhou}\corref{cor1}}
\ead{bo.zhou@yale.edu}
\author[1]{Huidong \snm{Xie}}
\author[1]{Qiong \snm{Liu}}
\author[1]{Xiongchao \snm{Chen}}
\author[1]{Xueqi \snm{Guo}}
\author[4]{Zhicheng \snm{Feng}}
\author[5]{Jun \snm{Hou}}
\author[6]{S. Kevin \snm{Zhou}}
\author[7]{Biao \snm{Li}}
\author[8]{Axel \snm{Rominger}}
\author[8,9]{Kuangyu \snm{Shi}}
\author[1,2,3]{James S. \snm{Duncan}}
\author[1,2]{Chi \snm{Liu}\corref{cor1}}
\cortext[cor1]{Corresponding author.}
\ead{chi.liu@yale.edu}

\address[1]{Department of Biomedical Engineering, Yale University, New Haven, CT, USA.}
\address[2]{Department of Radiology and Biomedical Imaging, Yale School of Medicine, New
Haven, CT, USA}
\address[3]{Department of Electrical Engineering, Yale University, New Haven, CT, USA.}
\address[4]{Department of Biomedical Engineering, University of Southern California, Los Angeles, CA, USA}
\address[5]{Department of Computer Science, University of California Irvine, Irvine, CA, USA}
\address[6]{School of Biomedical Engineering \& Suzhou Institute for Advanced Research, University of Science and Technology of China, Suzhou, China}
\address[7]{Department of Nuclear Medicine, Ruijin Hospital, Shanghai Jiao Tong University School of Medicine, Shanghai, China}
\address[8]{Department of Nuclear Medicine, Inselspital, Bern
University Hospital, University of Bern, Bern, Switzerland}
\address[9]{Computer Aided Medical Procedures and Augmented Reality, Institute of Informatics I16, Technical University of Munich, Munich, Germany}

% \title{Type the title of your paper, only capitalize first
% word and proper nouns\tnoteref{tnote1}}%
% \tnotetext[tnote1]{This is an example for title footnote coding.}

% \author[1]{Given-name1 \snm{Surname1}\corref{cor1}}
% \cortext[cor1]{Corresponding author: 
%   Tel.: +0-000-000-0000;  
%   fax: +0-000-000-0000;}
% \author[1]{Given-name2 \snm{Surname2}\fnref{fn1}}
% \fntext[fn1]{This is author footnote for second author.}
% \author[2]{Given-name3 \snm{Surname3}}
% %% Third author's email
% \ead{author3@author.com}
% \author[2]{Given-name4 \snm{Surname4}}

% \address[1]{Affiliation 1, Address, City and Postal Code, Country}
% \address[2]{Affiliation 2, Address, City and Postal Code, Country}

% \received{1 May 2013}
% \finalform{10 May 2013}
% \accepted{13 May 2013}
% \availableonline{15 May 2013}
% \communicated{S. Sarkar}

\begin{abstract}
Low-count PET is an efficient way to reduce radiation exposure and acquisition time, but the reconstructed images often suffer from low signal-to-noise ratio (SNR), thus affecting diagnosis and other downstream tasks. Recent advances in deep learning have shown great potential in improving low-count PET image quality, but acquiring a large, centralized, and diverse dataset from multiple institutions for training a robust model is difficult due to privacy and security concerns of patient data. Moreover, low-count PET data at different institutions may have different data distribution, thus requiring personalized models. While previous federated learning (FL) algorithms enable multi-institution collaborative training without the need of aggregating local data, addressing the large domain shift in the application of multi-institutional low-count PET denoising remains a challenge and is still highly under-explored. In this work, we propose FedFTN, a personalized federated learning strategy that addresses these challenges. FedFTN uses a local deep feature transformation network (FTN) to modulate the feature outputs of a globally shared denoising network, enabling personalized low-count PET denoising for each institution. During the federated learning process, only the denoising network's weights are communicated and aggregated, while the FTN remains at the local institutions for feature transformation. We evaluated our method using a large-scale dataset of multi-institutional low-count PET imaging data from three medical centers located across three continents, and showed that FedFTN provides high-quality low-count PET images, outperforming previous baseline FL reconstruction methods across all low-count levels at all three institutions. 
\end{abstract}

\begin{keyword}
%% MSC codes here, in the form: \MSC code \sep code
%% or \MSC[2008] code \sep code (2000 is the default)
\MSC 41A05\sep 41A10\sep 65D05\sep 65D17
%% Keywords
\KWD low-count PET\sep Deep Reconstruction\sep Personalized Federated Learning\sep Denoising
\end{keyword}

\end{frontmatter}

%\linenumbers

%% main text
\section{Introduction}
Positron Emission Tomography (PET) is a commonly used functional imaging
modality with wide applications in oncology, cardiology, neurology, and biomedical research. To reconstruct high-quality PET, the patient is injected with a customized dose of radioactive tracer which inevitably introduces radiation exposure to both patients and healthcare providers. Adhere to the principle of As Low As Reasonably Achievable (ALARA), minimizing the radiation dose is of great interest to patients \citep{strauss2006alara}, particularly for  PET applications where serial scans are commonly required. However, reducing the injection dose in PET would result in increased image noise, poor signal-to-noise ratio (SNR) and image artifacts, which would jeopardize the downstream clinical tasks. 

To generate high-quality PET from low-count PET, deep learning-based low-count PET imaging methods have been extensively explored \citep{xiang2017deep,wang20183d,lu2019investigation,kaplan2019full,hu2020dpir,gong2020parameter,zhou2020supervised,ouyang2019ultra,chen2019ultra,liu2020noise,liu2021artificial,song2021noise2void,liu2021artificial,gong2021evolution}, which have demonstrated superior performance than conventional methods \citep{dutta2013non,maggioni2012nonlocal,mejia2016noise}. While deep learning-based methods achieve promising performance, they often rely on training using diverse and large-scale paired low-count and full-count datasets that are often prohibitively expensive and difficult to collect. Even though we can alleviate this issue through building a centralized large-scale dataset by transferring all institutional data, the concerns of medical data privacy and security, the difficulty of building data transfer and warehouse protocol, and the laborious process make it challenging to implement this solution in practice \citep{rieke2020future, roski2014creating}.

Federated learning (FL) has recently emerged as a solution to address data privacy concerns in training deep models. This approach enables different local clients to collaboratively learn using their own data and computing resources, without sharing any private data. A client-to-cloud platform is established where the cloud server periodically communicates with local clients to collect local models. These models are then aggregated to generate a global model that is redistributed to the local clients for further local updates. Unlike traditional methods where local data is required to be directly transferred for global training, FL only involves the exchange of model parameters or gradients. As a result, FL can potentially solve the data privacy concerns for training a global model. In the context of low-count PET imaging, the different institutions may have different low-count protocols, different PET systems from different vendors, different reconstructions, and different post-processing protocols, thus can lead to significant data heterogeneity and domain shifts. Unfortunately, it is challenging to generalize a global model trained from the classical FL algorithm, such as FedAvg \citep{mcmahan2017communication}, to different institutions due to different data distributions at each local site. Thus, personalized federated learning is desirable to address this issue. 

Previous works have attempted to address the domain shift issues, but mainly focus on image classification \citep{li2021fedbn,arivazhagan2019federated,collins2021exploiting,fallah2020personalized,shamsian2021personalized}. In the application of accelerated MR reconstruction using FL, \cite{guo2021multi} first proposed to address the domain shift issue by iteratively aligning the latent feature of UNet \citep{ronneberger2015u} between target and other client sites. However, their cross-site strategy requires the target client to share both the latent feature and the network parameter with other client sites in each communication round, which could result in additional data privacy concerns \citep{lyu2022privacy,huang2021evaluating}. Similarly, \cite{feng2022specificity} proposed a UNet with a globally shared encoder for generalized representation learning and a client-specific decoder for domain-specific reconstruction, but the network architecture is limited to UNet or its variants. While achieving promising performance for the MR reconstruction task, the network architecture is limited to UNet or its variants due to the constraint in their FL algorithm designs. Additionally, instead of using simple encoder-decoder structures for improving image quality, state-of-the-art deep learning-based image restoration networks often deploy advanced designs, such as original resolution restoration \citep{zhang2018residual,zhou2021limited}, recurrent restoration \citep{zhou2020dudornet,zhou2023dsformer}, and multi-stage restoration \citep{zamir2021multi,zhou2022dudoufnet}. Moreover, all the previous works are limited to 2D MRI reconstruction while PET requires 3D processing. In the application of low-count PET denoising using FL, \cite{zhou2022federated} proposed the first FL study for low-count PET reconstruction. However, their study was performed only on simulation experiments with heterogeneous low-count data generated from one scanner at one site. In addition, their generation of personalized models relied on local fine-tuning after global training from FedAvg\citep{mcmahan2017communication}, which may not be optimal. Therefore, the development of 3D personalized federated learning framework and conducting studies with real-world multi-institutional low-count PET imaging data is desirable.

To address these challenges, we propose a personalized federated learning method based on deep feature transformation networks (FedFTN) and perform studies on real-world multi-institutional low-count PET data collected from three medical institutions across three continents, with multiple low-count levels contained at each institution. The general idea is to use a Feature Transformation Network (FTN) to modulate the features in the denoising network. During federated training, the FTNs are kept local, while only the denoising networks' parameters at different institutions are shared and aggregated at the central server. The FTNs are only trained locally to modulate the features from the denoising network, thus enabling personalized PET denoising at each institution. The input to the FTN is the low-count level for individual patients, thus also allowing using a single unified model for multiple low-count levels' denoising at each institution. In addition, we propose a Global Weight Constraint (GWC) loss which helps stabilize the local weight updates of the denoising network during the federated reconstruction learning. Our experimental result on the real-world multi-institutional low-count PET datasets demonstrates that our FedFTN can generate superior low-count PET denoising results as compared to previous federated reconstruction learning methods, as well as locally trained models. 

\begin{figure*}[htb!]
\centering
\includegraphics[width=0.92\textwidth]{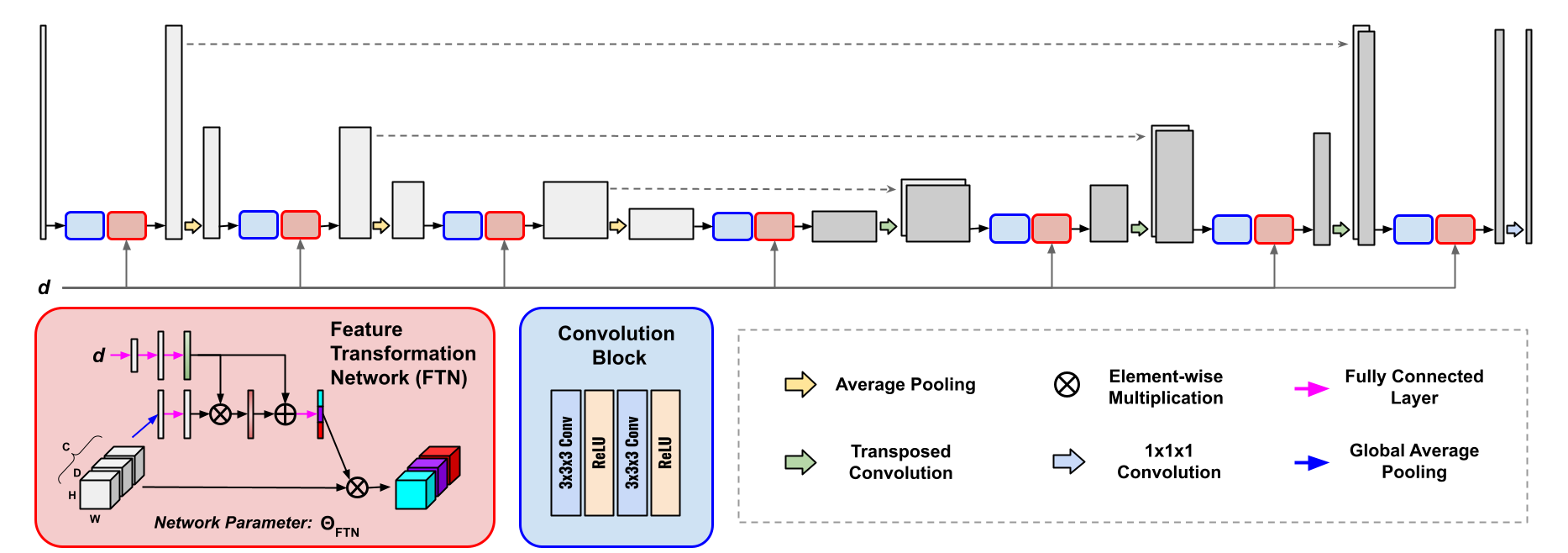}
\caption{The architecture of our Feature Transformation Network (FTN)-modulated Denoising Network. Without loss of generality, we deploy a U-Net as our denoising network. The FTNs are appended to the feature outputs at all U-Net resolution levels to transform the features at all levels.}
\label{fig:network}
\end{figure*}

% -------------------------------------------------------------------
\section{Related Work}
\noindent\textbf{Low-count PET Denoising.} Previous studies on low-count PET denoising can be divided into two categories: conventional post-processing methods \citep{dutta2013non,maggioni2012nonlocal,mejia2016noise} and deep learning-based methods \citep{xiang2017deep,wang20183d,lu2019investigation,kaplan2019full,hu2020dpir,gong2020parameter,zhou2020supervised,ouyang2019ultra,chen2019ultra,liu2020noise,liu2021artificial,song2021noise2void}. Although conventional methods like Gaussian filtering are standard post-processing techniques for PET reconstruction, they tend to oversmooth the image and have difficulty preserving local structures under low-count conditions with amplified noise. On the other hand, deep learning-based denoising methods have been extensively explored for low-count PET and have demonstrated promising results. For example, \cite{kaplan2019full} proposed to use a 2D Generative Adversarial Network (GAN) with UNet as a generator to predict full-count PET images from low-count PET images. \cite{wang20183d} proposed the use of a 3D conditional GAN to directly translate 3D low-count PET images to full-count PET images. Similarly, \cite{gong2020parameter} also proposed to use a 3D Wasserstein GAN to stabilize GAN training and to improve the low-count PET denoising performance. Based on previous GAN designs, \cite{ouyang2019ultra} further proposed to reinforce the denoising performance by incorporating patient-specific information. In parallel to using only the low-count PET images as network input, low-count PET denoising facilitated by other imaging modalities has also been explored. For example, \cite{xiang2017deep} developed an auto-context CNN using both low-count PET images and T1 MR images as inputs for full-count PET generation. Similarly, \cite{chen2019ultra} proposed to input low-count PET images along with multi-contrast MR images into a UNet for ultra-low-count PET denoising. There were also recent developments on unifying the low-count PET denoising model for multiple low-count levels \citep{xie2023unified}. In addition to these previous studies on static low-count PET, methods have also been developed for simultaneous motion correction and low-count PET reconstruction \citep{zhou2021mdpet,zhou2020simultaneous,zhou2023fast}, and have demonstrated further improved reconstruction quality. Even though deep learning-based methods have shown great promise in improving low-count PET images, these methods have so far been studied only within single institutions, where a single low-count protocol, identical PET system, and reconstruction protocol are assumed. Investigation on how to train generalizable reconstruction models using multi-institutional low-count data with non-identical distribution while addressing data privacy issues is of great significance and remains relatively unexplored.

\noindent\textbf{Federated Learning for Reconstruction.} FL with a decentralized learning framework enables multiple local institutions to collaborate in training shared models while maintaining their local data privacy \citep{li2020federated}. However, traditional FL algorithms, such as FedAvg \citep{mcmahan2017communication}, do not adequately address the domain shift issue, which arises due to differences in data distribution across clients despite allowing collaborative training without sharing data. To overcome this limitation, recent studies have proposed personalized federated learning strategies in medical image reconstruction tasks. For example, \cite{guo2021multi} proposed federated learning with cross-site modeling (FLCM) to address domain shift issues in 2D MR reconstruction by iteratively aligning the latent feature distribution between clients. However, FLCM's requirement for frequent communication between clients and latent vectors increases the communication cost and risk of potential privacy leakage. On the other hand, \cite{feng2022specificity} proposed using a client-specific decoder in a UNet to reduce domain shift while only uploading and aggregating the encoder's parameters. To further enable flexible MRI accelerated imaging operator across sites and improve generalizability, \cite{elmas2022federated} proposed FedGIMP that aims to generate site-specific MRI prior which uses a mapper network to produce site-specific latents for the generative network given a site index. A similar strategy has been also adapted to multi-contrast MRI synthesis \citep{dalmaz2022one}. Although these methods show promising performance, they have only been tested on 2D MRI reconstruction tasks, and many rely on specific network architecture, e.g. UNet or Auto-encoder, with latent representation for domain alignment and personalization. They may not generalize well to applications requiring other network architectures, such as cascade/recurrent/multistage network designs \citep{zhou2020dudornet,zhou2021limited,zhou2022dudodr,zhou2022dudoufnet}, for better reconstruction quality. For CT reconstruction using FL, strategies have also been developed to adapt to different CT distributions from different sparse-view CT (SVCT) protocols. For example, with simulated multi-institutional 2D CT data, \cite{yang2022hypernetwork} proposed to extend the PDF framework \citep{xia2021ct} into a hypernet-based FL framework, where the normalization parameter generator sub-networks were kept locally to adapt to different SVCT distributions. For low-count PET denoising, \cite{zhou2022federated} proposed the first FL study for low-count PET imaging. However, their study was limited to simulation experiments with heterogeneous low-count data generated from one scanner at one site. Furthermore, their approach of generating personalized models relied on local fine-tuning after global training from FedAvg\citep{mcmahan2017communication}, which may not be optimal. Therefore, there is a need to develop a 3D personalized federated learning approach and conduct studies with real-world multi-institutional low-count PET data.

\begin{figure*}[htb!]
\centering
\includegraphics[width=0.96\textwidth]{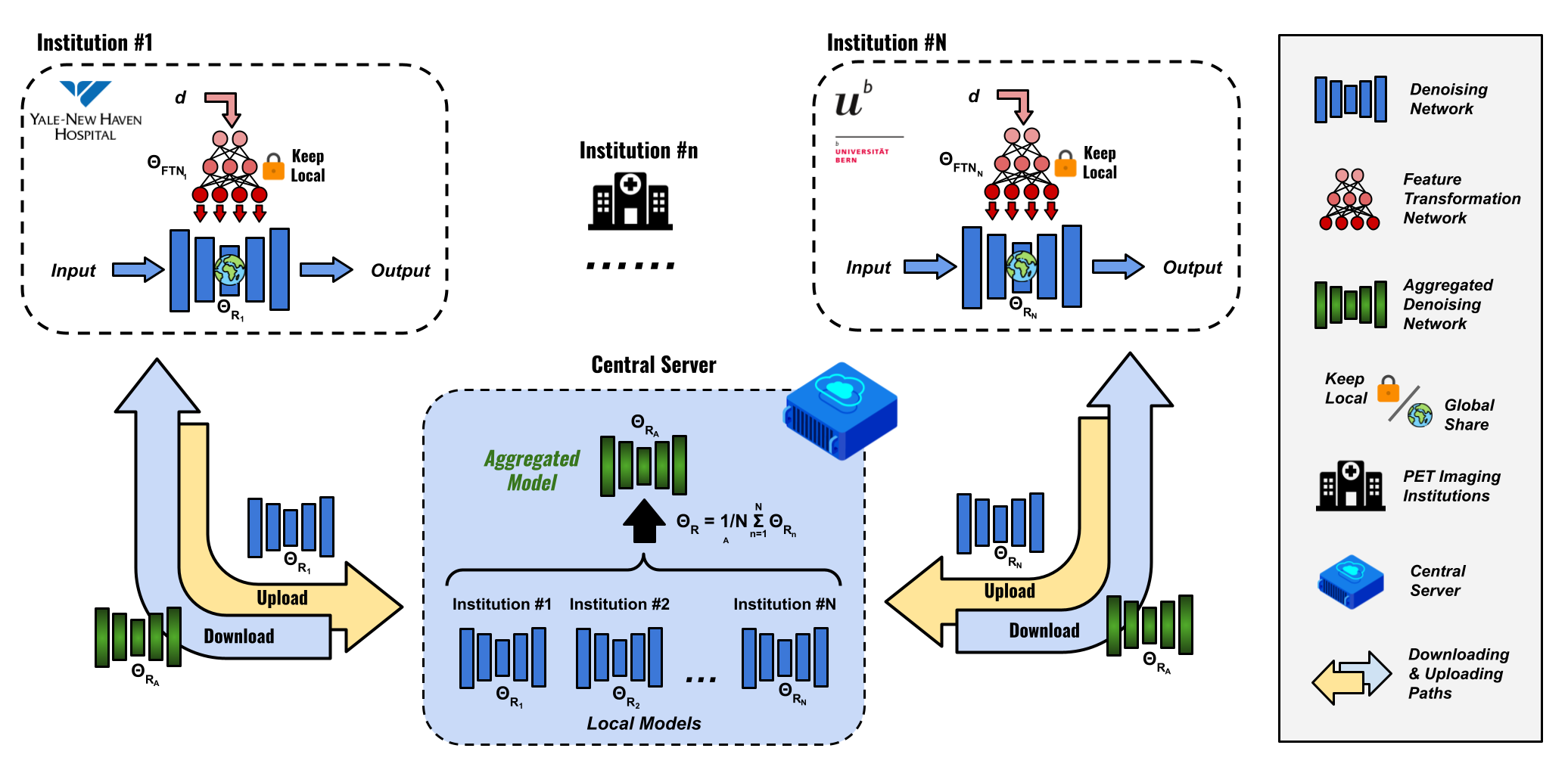}
\caption{Personalized federated learning based on feature transformation network (FedFTN). Each institution contains one FTN-modulated denoising network (Figure \ref{fig:network}). At each global iteration, each FTN-modulated denoising network at each institution is first trained locally for a fixed number of epochs. During model aggregation, the FTNs are kept local and only the denoising networks' parameters are uploaded to the central cloud server for parameter averaging. The algorithm is summarized in Algorithm \ref{alg:train_alg}. }
\label{fig:framework}
\end{figure*}

% -------------------------------------------------------------------
\section{Methods}
Our method included a Feature Transformation Network (FTN)-modulated denoising network (Figure \ref{fig:network}) and a customized personalized federated learning framework (Figure \ref{fig:framework}) for it. Details were elaborated in the following sections.

\subsection{Feature Transformation Network}
The Feature Transformation Network (FTN) aimed to modulate the global-sharing denoising network to generate personalized denoised images that are specific to the local site and the target low-count level within the site. 

As illustrated in Figure \ref{fig:network}, given the extracted intermediate features from the convolution blocks of the global-sharing denoising network, the features were inputted into the FTNs for personalized adaptation. Specifically, given a feature input $F = [f_1, f_2, \dots, f_C]$ with $f_n \in \mathbb{R}^{H \times W \times D}$ denoting the individual feature channel, we flattened the feature via global average pooling, generating vector $v \in \mathbb{R}^{C}$ with its $z$-th element:
\begin{equation}
    v_z = \frac{1}{H \times W \times D} \sum^H_i \sum^W_j \sum^D_k f_z (i,j,k) , 
\end{equation}
where vector $v$ embedded the global information of the input feature. Then, $v$ was fed into a fully connected layer with weights of $w_R \in \mathbb{R}^{C \times C}$ and generated $v_R = w_R v$.

In parallel, given the low-count levels $d \in \mathbb{R}^{1}$ at the local site, we deployed three consecutive fully connected layers with weights of $w_{1} \in \mathbb{R}^{\frac{C}{2} \times 1}$, $w_{2} \in \mathbb{R}^{C \times \frac{C}{2}}$, and $w_{3} \in \mathbb{R}^{C \times C}$, which generated transformation guiding vector:
\begin{equation}
    v_d = w_{3} \eta(w_{2} \eta(w_{1} d)) , 
\end{equation}
where $\eta$ is the ReLU activation function. Then, $v_R$ and $v_d$ were fused via:
\begin{equation}
    v_{fuse} = \sigma (v_d) v_R + v_d  , 
\end{equation}
where $\sigma$ is the Sigmoid function. Then, we used another fully connected layer with weights of $w_{fuse} \in \mathbb{R}^{C \times C}$ to generate the transformation vector:
\begin{equation}
    \hat{v} = w_{fuse} v_{fuse} 
\end{equation}
Finally, the transformation vector was applied to the input feature map using channel-wise multiplication:
\begin{equation}
    F_{out} = [f_1 \hat{v}_1, f_2 \hat{v}_2, \dots, f_C \hat{v}_C]
\end{equation}

Please note the FTNs were deployed at all feature resolution levels in the denoising network (Figure \ref{fig:network}), thus allowing progressive feature transformation for personalized denoising. 

\begin{algorithm2e*}[!htb]
\caption{Steps of our Personalized Federated Learning Framework.}\label{alg:train_alg} 

\textbf{Input:} $\mathcal{D}$ = \{$\mathcal{D}_1$, $\mathcal{D}_2$, \ldots, $\mathcal{D}_N$\} \Comment*[r]{ Training datasets from N institutions}

\textbf{Initialize \#1:} {P: the number of local epochs; Q: the number of global epochs}

\textbf{Initialize \#2:} {$\theta_{R}$: parameters of global denoising network}

\textbf{Initialize \#3:} {$\theta_{FTN_1}$, \ldots, $\theta_{FTN_N}$: parameters of local feature transformation networks}

\textbf{Initialize \#4:} {$\alpha$: local training learning rate; $\lambda$: weight of GWC loss}

\For{$q = 1$ to $Q$}
{
	\For{$n = 1$ to $N$ in parallel}
    {
        $\rightarrow$ deploy averaged global denoising network parameters $\theta_{R_A}$ to local institution, if available
        
        \For{$p = 1$ to $P$}
        {
        \If{$q < 3$}
        {
         $\mathcal{L}_n = \sum\limits_{(x,y,d) \in \mathcal{D}_n} ||G_n(x, d | \theta_{R_n}, \theta_{FTN_n}) - y||_2^2 $  \Comment*[r]{Compute local training loss (warm up)}
         }
         \Else{
         $\mathcal{L}_n = \sum\limits_{(x,y,d) \in \mathcal{D}_n} ||G_n(x, d | \theta_{R_n}, \theta_{FTN_n}) - y||_2^2 + \lambda || \theta_{R_n} - \theta_{R_A} ||_2^2 $  \Comment*[r]{Compute local training loss}
         }
         
         $\{\theta_{R_n}^{p+1}, \theta_{FTN_n}^{p+1}\}  \leftarrow \{\theta_{R_n}^{p}, \theta_{FTN_n}^{p}\} - \alpha \nabla \mathcal{L}_n$   \Comment*[r]{Update network parameters}
        }
        
        $\rightarrow$ upload global denoising network parameters $\theta_{R_n}^q$ to the central server
        
    }
    
$\theta_{R_A}^q = \frac{1}{N} \sum\limits_{n=1}^{N} \theta_{R_n}^q$  \Comment*[r]{Update global-sharing denoising network parameter by averaging}

}
\textbf{Output} $\{\theta_{R_A}^Q, \theta_{FTN_1}^Q\}$, $\{\theta_{R_A}^Q, \theta_{FTN_2}^Q\}$, \ldots, $\{\theta_{R_A}^Q, \theta_{FTN_N}^Q\}$  \Comment*[r]{Return personalized models for each institution}

\end{algorithm2e*} 

\subsection{FTN-based Personalized Federated Learning}
The general pipeline of the FTN-based personalized federated learning is shown in Figure \ref{fig:framework}. We denoted $\mathcal{D}_1, \mathcal{D}_2, \ldots, \mathcal{D}_N$ as the low-count PET datasets from $N$ different institutions. Each institutional dataset $\mathcal{D}_n$ contained pairs of full-count and low-count PET 3D images, where each local dataset has three different low-count level settings. Within each institution, an FTN-modulated denoising network can be trained using the local data with two loss components, including a restoration loss $\mathcal{L}_{recon}$ and a global weight constraint (GWC) loss $\mathcal{L}_{gwc}$. The first loss can be formulated as:
\begin{equation}\label{eq:loss}
    \mathcal{L}_{re} = \sum\limits_{(x,y,d) \in \mathcal{D}_n} ||G_n(x,d|\theta_{R_n}, \theta_{FTN_n}) - y||_2^2 , 
\end{equation}
where $G_n$ is the FTN-modulated denoising network at the $n^{th}$ institution, and was parameterized by $\theta_{R_n}$ and $\theta_{FTN_n}$. $\theta_{R_n}$ is the denoising network's parameters, and $\theta_{FTN_n}$ is the FTN's parameters. $x$, $y$, and $d$ are the training pair of the low-count image, full-count image, and low-count level from $\mathcal{D}_n$. We assume there were $Q$ global training epochs (communication between cloud server \& local institutions), and $P$ local training epochs at each institution. During the first two global epochs, the iterative optimization of the local network's parameter only used the reconstruction loss and can be written as:
\begin{equation}
    \{\theta_{R_n}^{p+1}, \theta_{FTN_n}^{p+1}\}  \leftarrow \{\theta_{R_n}^{p}, \theta_{FTN_n}^{p}\} - \alpha \nabla \mathcal{L}_{recon}
\end{equation}
where $\alpha$ is the learning rate. At the end of each global training epoch, the global-sharing denoising network's parameters $\theta_{R_n}$ at each institution can be uploaded to a cloud server for weight aggregation. The cloud server updated the parameters of the global-sharing denoising network by:
\begin{equation}
    \theta_{R_A}^q = \frac{1}{N} \sum\limits_{n=1}^{N} \theta_{R_n}^q
\end{equation}
where $q$ denotes the $q^{th}$ global epoch. After two global epochs of warm-up, we further added a GWC loss to stabilize the weights update during the local training in the following global epochs. The GWC loss is formulated as:
\begin{equation}
    \mathcal{L}_{gwc} = || \theta_{R_n} - \theta_{R_A} ||_2^2 
\end{equation}
where the current denoising network's weight $\theta_{R_n}$ was constrained to be in proximal to the aggregated weight $\theta_{R_A}$. Starting from the 3rd global epoch, the training loss was the combination of the reconstruction loss and the global weight constraint loss, thus can be written as:
\begin{equation}
    \mathcal{L}_{comb} = \mathcal{L}_{re} + \lambda \mathcal{L}_{gwc}
\end{equation}
where $\lambda$ is the weight of GWC loss and we set it to $0.001$. Then, the iterative optimization of the local network's parameter can be formulated as:
\begin{equation}
    \{\theta_{R_n}^{p+1}, \theta_{FTN_n}^{p+1}\}  \leftarrow \{\theta_{R_n}^{p}, \theta_{FTN_n}^{p}\} - \alpha \nabla \mathcal{L}_{comb}
\end{equation}

After $Q$ rounds of communication between local institutions and the cloud server, we can obtain collaboratively-trained FTN-modulated denoising networks with personalized parameters $\{\theta_{R_A}^Q, \theta_{FTN_1}^Q\}$, $\{\theta_{R_A}^Q, \theta_{FTN_2}^Q\}$, \ldots, $\{\theta_{R_A}^Q, \theta_{FTN_N}^Q\}$. The algorithm is summarized in Algorithm \ref{alg:train_alg}. 

\subsection{Multi-institutional low-count PET Data}
We collected multi-institutional low-count PET data from three different medical centers in USA, Switzerland, and China for our study. The first dataset was collected at Yale New Haven Hospital, New Haven, USA. 200 subjects were included in this dataset. The subjects were injected with a \textsuperscript{18}F-FDG tracer and the whole-body protocol with continuous-bed motion scanning was used. All data were acquired using a Siemens Biograph mCT PET/CT system. The average dose across all patients is $256.3\pm16.2$MBq. We used uniform down-sampling of the PET list-mode data with down-sampling ratios of 5\%, 10\%, and 20\% to generate low-count PET data at three different low-count levels. For both the low-count and full-count images, they were reconstructed using the ordered-subsets expectation maximization (OSEM) algorithm with 2 iterations and 21 subsets, provided by the vendor. A post-reconstruction Gaussian filter with $5 mm$ full width at half maximum (FWHM) was used. The voxel size of the reconstructed image was $2.04\times 2.04 \times 2.03 mm^3$. The image size was $400 \times 400$ in the transverse plane and varied in the axial direction depending on patient height. The 200 subjects were split into 100 subjects for training, 10 subjects for validation, and 90 subjects for evaluation. The second dataset was collected at the Department of Nuclear Medicine, University of Bern, Bern, Switzerland \citep{xue2021cross}. 209 subjects with \textsuperscript{18}F-FDG tracer were included in this dataset. All data were acquired using a Siemens Biograph Vision Quadra whole-body PET/CT system. The average dose across all patients is $264.1\pm18.2$MBq. Here, low-count PET data at 2\%, 5\%, and 10\% low-count levels were generated by down-sampling of the PET list-mode data. For both the low-count and full-count images, they were reconstructed using the OSEM algorithm with 6 iterations and 5 subsets, provided by the vendor. A post-reconstruction Gaussian filter with $5 mm$ FWHM was used. The voxel size of the reconstructed image was $1.65 \times 1.65 \times 1.65 mm^3$. The image size was $440 \times 440 \times 644$. The 209 subjects were split into 109 subjects for training, 10 subjects for validation, and 90 subjects for evaluation. The third dataset was collected at the Ruijin Hospital, Shanghai, China \citep{xue2021cross}. 204 subjects with \textsuperscript{18}F-FDG tracer were included in this dataset. All data were acquired using a United Imaging uExplorer total-body PET/CT system. The average dose across all patients is $260.1\pm12.2$MBq Here, low-count PET data at 2\%, 5\%, and 10\% low-count levels were generated by down-sampling of the PET list-mode data. For both the low-count and full-count images, they were reconstructed using the OSEM algorithm with 4 iterations and 20 subsets, provided by the vendor. A post-reconstruction Gaussian filter with $5 mm$ FWHM was used. The voxel size of the reconstructed image was $1.66 \times 1.66 \times 2.88 mm^3$. The image size was $360 \times 360 \times 674$. The 204 subjects were split into 104 subjects for training, 10 subjects for validation, and 90 subjects for evaluation.

\subsection{Evaluation Metrics and Baselines Comparisons}
We evaluated the low-count denoised results using Peak Signal-to-Noise Ratio (PSNR) and Normalized Mean Square Error (NMSE) computed against full-count reconstruction ground truth. For baseline comparisons, we first compared our results against previous federated reconstruction algorithms, including Federated Averaging (FedAvg, \cite{mcmahan2017communication}), Federated Learning with Proximal Term (FedProx, \cite{li2020federated}), Federated Learning with Local Batch Normalization (FedBN, \cite{li2021fedbn}), Specificity-Preserving Federated Learning (FedSP, \cite{feng2022specificity}), and Personalized Federated Learning with Hypernetwork (FedHyper, \cite{shamsian2021personalized}). For a fair comparison, all the methods used an identical reconstruction/denoising network, as shown in Figure \ref{fig:network}. We also compared our FedFTN against two types of locally trained models, including local single models and local unified models. For each local single model, an FTN-modulated denoising network was trained using one single low-count level data at one specific institution. On the other hand, the local unified models are also based on FTN-modulated denoising networks, but all three low-count levels' data at a specific institution are used as training data. Furthermore, we also performed ablative studies on federated transfer learning \citep{zhou2022federated}, where FedFTN are further fine-tuned using local data for site adaption. 

\subsection{Implementation Details}
We implemented our method in Pytorch and performed experiments using an NVIDIA Quadro RTX 8000 GPU with 48GB memory. The Adam solver was used to optimize our models with $lr = 1 \times 10^{-4}$, $\beta_{1} = 0.9$, and $\beta_{2} = 0.999$. We used a batch size of $3$ and trained all models for 300 global epochs. The number of the local epoch was set to $3$. To prevent overfitting at each local site, we also implemented ’on-the-fly’ data augmentation. During training, we performed 64 × 64 × 64 random cropping, and then randomly flipped the cropped volumes along the x, y, and z-axis. During the site adaptation with fine tuning on local data, we used a reduced learning rate of $2e-5$ and trained the FedFTN models using the local dataset for $10$ epochs, and the batch size was also set to $3$ with ’on-the-fly’ data augmentation. Our method's training takes about 160 hours to complete. For baseline methods, FedHyper takes about 172 hours, and the rest baselines also take about 160 hours.

\begin{figure}[htb!]
\centering
\includegraphics[width=0.46\textwidth]{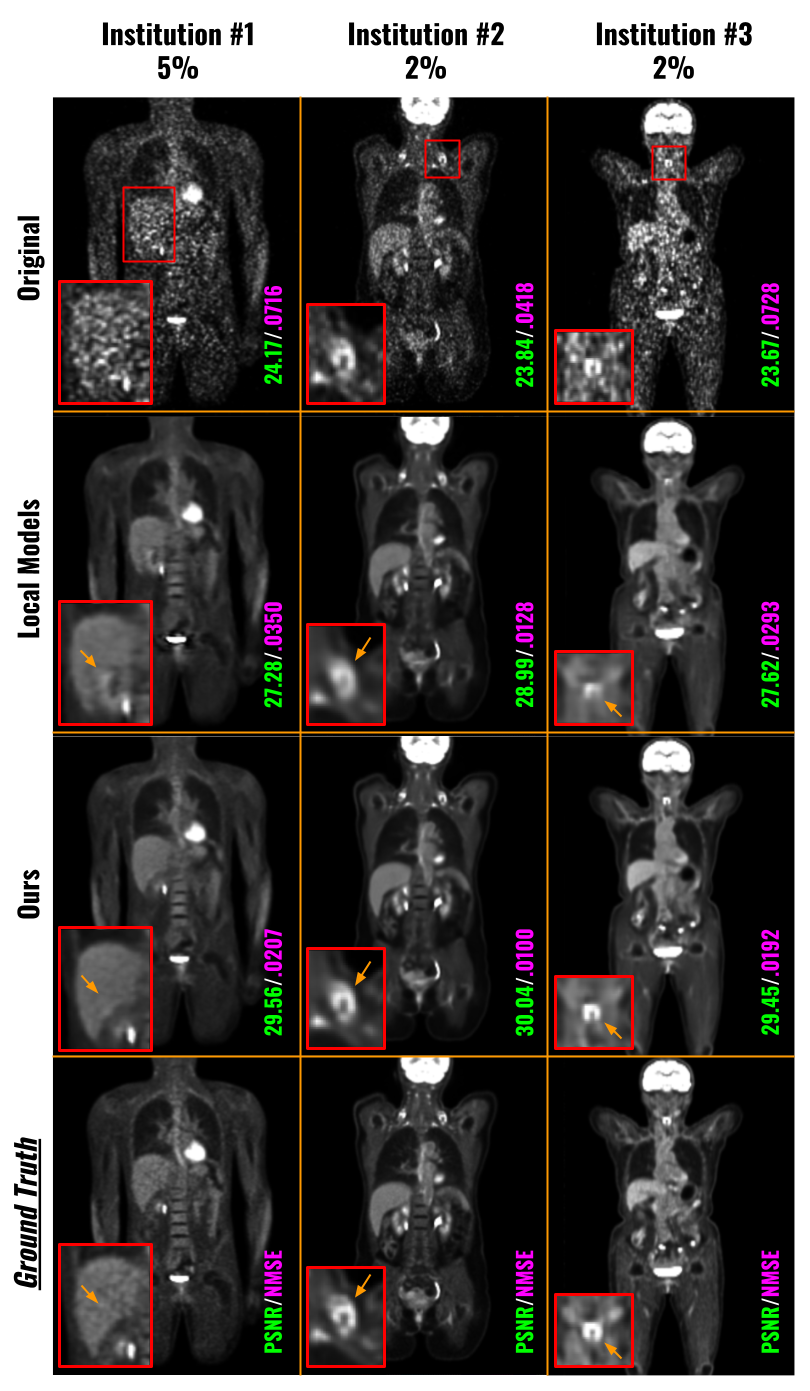}
\caption{Visual comparison of the low-count denoised images from locally trained models and from our FedFTN. The local single model means FTN-modulated denoising networks trained at one specific low-count level at the specific institution. The local unified model means FTN-modulated denoising networks trained with all three low-count levels within each institution. The PET images with the lowest count levels at each institution are visualized here.}
\label{fig:compare_local}
\end{figure}

\begin{table*} [htb!]
% \small
\footnotesize
\centering
\caption{Quantitative comparisons of the low-count PET denoised images between our FedFTN and locally trained models. The local single model means FTN-modulated denoising networks trained at one specific low-count level at the specific institution. The local unified model means FTN-modulated denoising networks trained with all three low-count levels within each institution. The performance of FedFTN with Site Adaptation (SA) via further local data fine-tuning is reported in the last row. The best results are marked in \textbf{bold}. "$\dagger$" indicates that the difference between FedFTN and all compared methods is significant at $p < 0.005$ based on the non-parametric Wilcoxon signed rank test.}
\label{tab:comp_local}
\resizebox{\textwidth}{!}{
    \begin{tabular}{l|c|c|c||c|c|c||c|c|c}
        \hline
        \textbf{Evaluation}      & \multicolumn{3}{c}{\textbf{Institution \#1}}    &  \multicolumn{3}{c}{\textbf{Institution \#2}}    &  \multicolumn{3}{c}{\textbf{Institution \#3}}                        \Tstrut\Bstrut\\
        \cline{2-10}
        \textbf{PSNR/NMSE/SSIM}       & 5\%    & 10\%    & 20\%    & 2\%    & 5\%    & 10\%    & 2\%    & 5\%    & 10\%                      \Tstrut\Bstrut\\
        \hline
        Original                 & $20.46/.129/.931$  & $23.80/.059/.954$  & $27.40/.026/.972$    & $19.90/.076/.923$  & $23.66/.031/.951$  & $26.26/.017/.968$   & $20.61/.121/.933$  & $24.89/.050/.961$  & $27.55/.030/.975$      \Tstrut\Bstrut\\
        Local Single Models      & $25.93/.034/.966$  & $27.91/.021/.975$  & $30.22/.012/.983$    & $25.37/.021/.971$  & $27.16/.014/.980$  & $28.64/.010/.985$   & $26.00/.035/.974$  & $28.17/.023/.982$  & $29.90/.018/.986$      \Tstrut\Bstrut\\
        Local Unified Model      & $26.02/.032/.968$  & $28.01/.020/.975$  & $30.21/.012/.983$    & $25.42/.020/.970$  & $27.21/.014/.980$  & $28.64/.010/.985$   & $26.02/.035/.975$  & $28.20/.023/.982$  & $29.91/.018/.986$      \Tstrut\Bstrut\\
        \hline
        FedFTN                   & $\mathbf{27.24/.025/.999}$$^\dagger$  & $\mathbf{28.96/.017/.999}$$^\dagger$  & $\mathbf{30.82/.011/.999}$$^\dagger$    & $\mathbf{26.12/.018/.999}$$^\dagger$  & $\mathbf{27.80/.013/.999}$$^\dagger$  & $\mathbf{29.03/.009/.999}$$^\dagger$     & $\mathbf{26.83/.031/.999}$$^\dagger$  & $\mathbf{28.91/.021/.999}$$^\dagger$  & $\mathbf{30.23/.017/.999}$$^\dagger$    \Tstrut\Bstrut\\
        \hline
    \end{tabular}
    }
\end{table*}

% -------------------------------------------------------------------
\section{Experimental Results}
Figure \ref{fig:compare_local} shows qualitative comparisons of low-count PET denoised images using locally trained denoising models and our FedFTN method. The lowest low-count level denoised results from each institution were visualized for comparison. As we can see, since only less than 5\% count was used, the original images (1st row) suffer from high noise and image artifacts. While the locally trained models can reduce the noise and recover the general structure, the detail recovery was still not ideal. For example, the 5\% low-count denoised image from the locally trained model at Institution \#1 created additional artifacts at the intersection between the liver and kidney. The 2\% low-count denoised images from the locally trained models at Institution \#2 and Institution \#3 suffered from heavy blurring on important regions, such as the hypermetabolic lesions in the zoomed boxes. The quantitative comparisons were summarized in Table \ref{tab:comp_local}. We reported quantitative results for all available low-count levels for the three institutions. Similar to the observation from the visualizations, all the original PET images suffered from low SNR, resulting in low PSNR and NMSE values. Taking Institution \#1 as an example, we can see the locally trained unified model, i.e. the FTN-modulated Denoising Network, was able to improve the PSNR from 20.46 to 26.02 for the 5\% low-count PET. Please note that the local unified model is the network shown in Figure \ref{fig:network} that is trained locally with the local site's low-count level as an additional input. Using the FedFTN, we can further improve the PSNR from 26.02 to 27.24 with statistical significance. Similar observations on improvement over the locally trained models can be found for other low-count levels at Institution \#1 and other institutions. The average inference time for testing data of institution \#1, institution \#2, and institution \#3 were $8.08 \pm 2.83$ s, $11.32 \pm 1.05$ s, and $10.04 \pm 1.09$ s, respectively.

\begin{figure*}[htb!]
\centering
\includegraphics[width=1.00\textwidth]{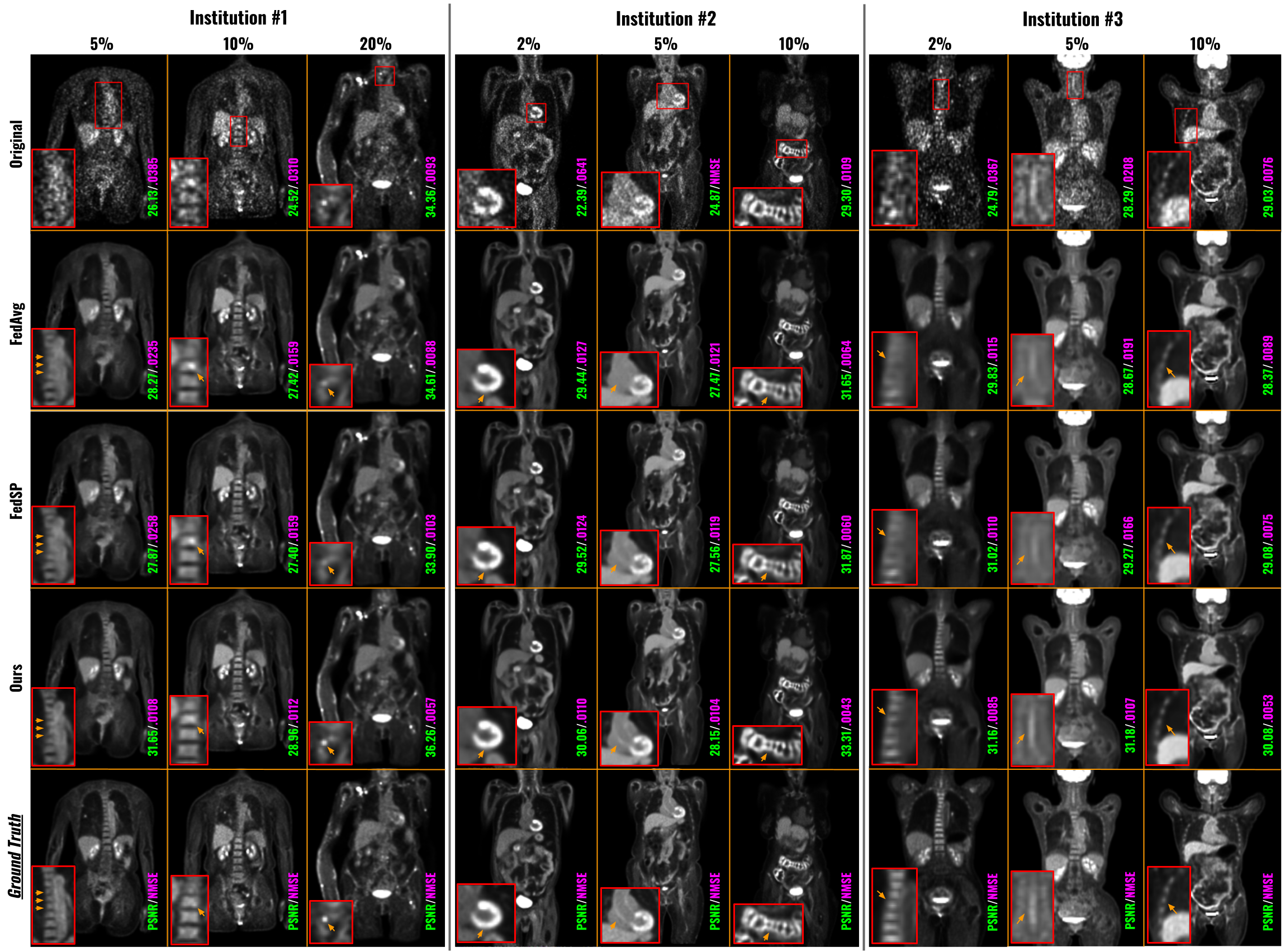}
\caption{Qualitative comparison of low-count denoising with different federated learning methods. Each institution contains three-different low-count levels. The original low-count reconstruction and the full-count ground truths are shown in the first row and last row, respectively. The image quality metrics of each image are indicated at the bottom of the images.}
\label{fig:compare_methods}
\end{figure*}

Figure \ref{fig:compare_methods} presented qualitative comparisons of low-count PET denoised images using different federated methods. For each institution, we showed one patient example for each low-count level, and all low-count levels are visualized. At Institution \#1, all three low-count levels' denoised images suffered from low SNR, with the noise level increasing as the dose level decreases. While previous FL methods can improve the image quality by suppressing the noise, the detail recovery was still suboptimal. For instance, for the 5\% low-count level, the bone marrows in the spine were heavily blurred by the previous methods, whereas FedFTN showed much sharper spine, demonstrating the best consistency with the ground truth full-count reconstruction. Additionally, for the 10\% low-count level, a strong false positive signal was visible in the spine from the original low-count reconstruction, and previous methods failed to suppress it, which may lead to misdiagnosis. In contrast, using FedFTN, we can suppress this false positive signal and provided a high-quality image that best matches the ground truth. At Institution \#2, for the 2\% low-count level, a false positive of cardiac defect was visible from the original low-count reconstruction. While previous methods can suppress the noise, this false positive defect signal cannot be fully removed, particularly for FedSP. In contrast, FedFTN demonstrated significantly better performance in false positive defect removal and the most consistent cardiac shape as compared to the ground truth. Similarly, for the 5\% low-count level, FedFTN better recovered the signal of the aorta wall than previous methods. We observe similar results in the patient examples from Institution \#3, where FedFTN can provide better image quality as compared to previous methods. The corresponding quantitative comparisons were summarized in Table \ref{tab:comp_methods}. As mentioned previously, the original low-count reconstructions from all three institutions suffered from poor image quality, where the PSNR values were all lower than $21.00$ at the lowest low-count levels for all three institutions. By deploying previous federate learning methods, as compared to the locally trained models (Figure \ref{fig:compare_local} and Table \ref{tab:comp_local}), we can see that these previous FL methods can already improve the low-count image quality across all institutions at all low-count levels. For example, FedSP was able to increase the PSNR from $26.02$ to $26.69$ for the 5\% low-count reconstruction at Institution \#1, from $25.42$ to $25.80$ for the 2\% low-count reconstruction at Institution \#2, and from $26.02$ to $26.44$ for the 2\% low-count reconstruction at Institution \#3. In the second last row, we showed that our FedFTN can significantly outperform these previous FL baselines, and achieved state-of-the-art performance across all the low-count levels at all three institutions. For example, as compared to the FedHyper with the best performance among previous FL reconstruction algorithms, our FedFTN can further improve the PSNR from $26.88$ to $27.24$ for the 5\% low-count reconstruction at Institution \#1, from $25.85$ to $26.12$ for the 2\% low-count reconstruction at Institution \#2, and from $26.44$ to $26.89$ for the 2\% low-count reconstruction at Institution \#3.  

\begin{table*} [htb!]
% \small
\footnotesize
\centering
\caption{Quantitative comparisons of low-count PET denoised images using different federated learning methods. Each institution contains three different low-count levels. The best results are marked in \textbf{bold}. "$\dagger$" indicates that the difference between FedFTN and all compared methods is significant at $p < 0.005$ based on the non-parametric Wilcoxon signed rank test.}
\label{tab:comp_methods}
\resizebox{\textwidth}{!}{
    \begin{tabular}{l|c|c|c||c|c|c||c|c|c}
        \hline
        \textbf{Evaluation}      & \multicolumn{3}{c}{\textbf{Institution \#1}}    &  \multicolumn{3}{c}{\textbf{Institution \#2}}    &  \multicolumn{3}{c}{\textbf{Institution \#3}}                        \Tstrut\Bstrut\\
        \cline{2-10}
        \textbf{PSNR/NMSE/SSIM}  & 5\%    & 10\%    & 20\%    & 2\%    & 5\%    & 10\%    & 2\%    & 5\%    & 10\%                      \Tstrut\Bstrut\\
        \hline
        Original                 & $20.46/.129/.931$  & $23.80/.059/.954$  & $27.40/.026/.972$    & $19.90/.076/.923$  & $23.66/.031/.951$  & $26.26/.017/.968$   & $20.61/.121/.933$  & $24.89/.050/.961$  & $27.55/.030/.975$      \Tstrut\Bstrut\\
        FedAvg                   & $26.62/.029/.969$  & $28.30/.019/.976$  & $29.95/.013/.981$    & $25.70/.020/.971$  & $27.35/.014/.980$  & $28.43/.010/.985$   & $26.07/.035/.973$  & $28.07/.023/.982$  & $29.16/.019/.986$      \Tstrut\Bstrut\\
        FedBN                    & $26.68/.028/.970$  & $28.35/.019/.978$  & $30.05/.013/.982$    & $25.79/.019/.973$  & $27.47/.013/.981$  & $28.60/.010/.986$   & $26.31/.033/.975$  & $28.34/.022/.983$  & $29.38/.018/.987$      \Tstrut\Bstrut\\
        FedProx                  & $26.64/.028/.969$  & $28.32/.018/.977$  & $29.99/.013/.981$    & $25.75/.020/.972$  & $27.39/.014/.981$  & $28.50/.010/.985$   & $26.13/.034/.974$  & $28.17/.023/.983$  & $29.23/.019/.986$      \Tstrut\Bstrut\\
        FedSP                    & $26.69/.028/.969$  & $28.36/.019/.976$  & $30.16/.013/.982$    & $25.80/.019/.973$  & $27.51/.013/.981$  & $28.63/.010/.985$   & $26.23/.034/.974$  & $28.33/.022/.983$  & $29.48/.018/.987$      \Tstrut\Bstrut\\
        FedHyper                 & $26.88/.027/.971$  & $28.56/.018/.978$  & $30.33/.012/.983$    & $25.85/.019/.974$  & $27.49/.013/.982$  & $28.65/.010/.986$   & $26.44/.033/.976$  & $28.52/.022/.983$  & $29.73/.018/.987$      \Tstrut\Bstrut\\
        \hline
        FedFTN                   & $\mathbf{27.24/.025/.979}$$^\dagger$  & $\mathbf{28.96/.017/.983}$$^\dagger$  & $\mathbf{30.82/.011/.990}$$^\dagger$    & $\mathbf{26.12/.018/.980}$$^\dagger$  & $\mathbf{27.80/.013/.989}$$^\dagger$  & $\mathbf{29.03/.009/.991}$$^\dagger$     & $\mathbf{26.83/.031/.980}$$^\dagger$  & $\mathbf{28.91/.021/.990}$$^\dagger$  & $\mathbf{30.23/.017/.992}$$^\dagger$    \Tstrut\Bstrut\\
        \hline
        FedFTN + SA                   & $\mathbf{27.32/.024/.980}$$^\dagger$  & $\mathbf{28.99/.016/.985}$$^\dagger$  & $\mathbf{30.83/.011/.991}$$^\dagger$    & $\mathbf{26.21/.017/.981}$$^\dagger$  & $\mathbf{27.84/.012/.990}$$^\dagger$  & $\mathbf{29.05/.009/.992}$$^\dagger$    & $\mathbf{26.89/.031/.980}$$^\dagger$  & $\mathbf{28.94/.021/.991}$$^\dagger$  & $\mathbf{30.25/.017/.992}$$^\dagger$    \Tstrut\Bstrut\\
        \hline
    \end{tabular}
    }
\end{table*}

\begin{figure}[htb!]
\centering
\includegraphics[width=0.46\textwidth]{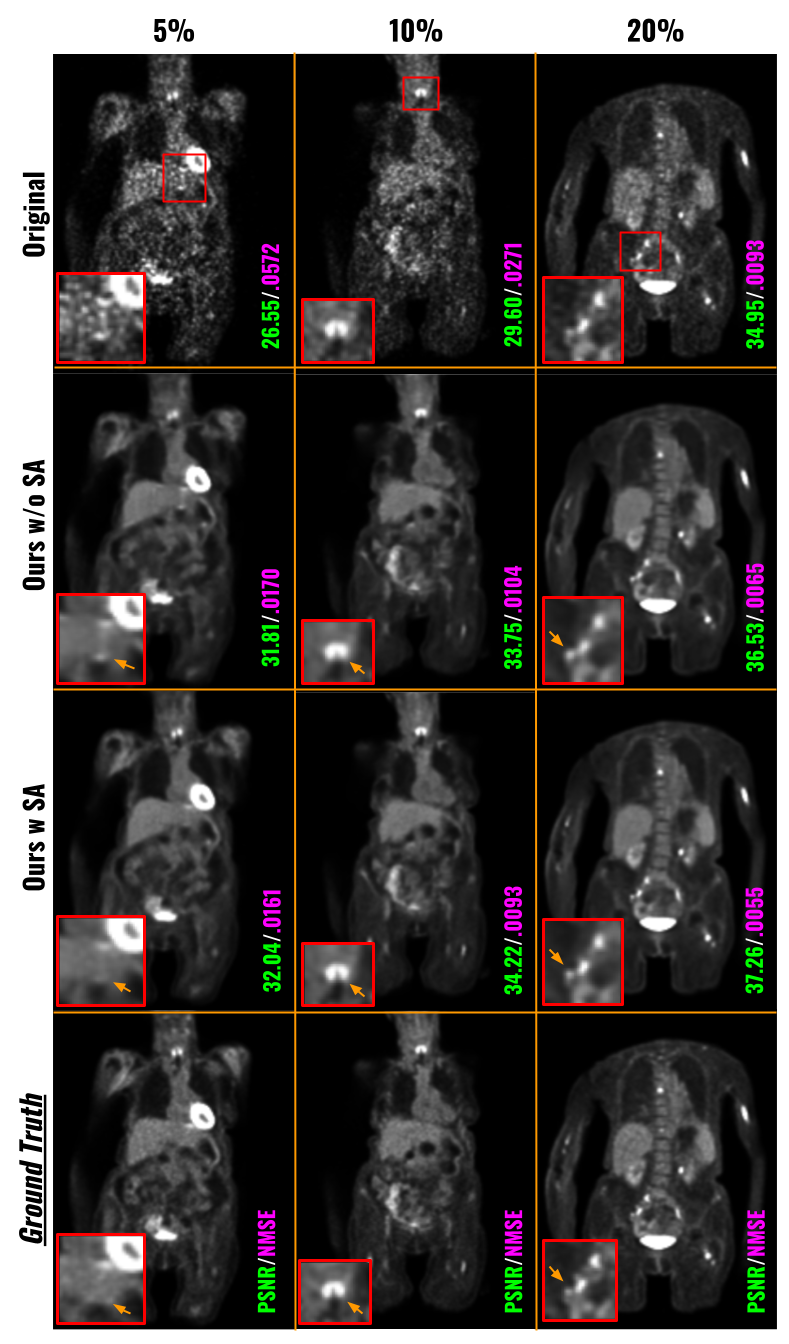}
\caption{Visual comparison of low-count PET denoised images from FedFTN before and after Site Adaptation (via) local fine-tuning. The three low-count levels at the Institution \#1 are shown here.}
\label{fig:compare_sa}
\end{figure}

Similar to the process of FTL \citep{zhou2022federated}, we further performed Site Adaptation (SA) through local fine-tuning for our FedFTN. The quantitative results were reported in the last row in Table \ref{tab:comp_methods}. We can observe further improved image quality metrics of FedFTN through SA for all low-count levels at all three institutions. In addition, we also found the denoising results from FedFTN with SA still significantly outperformed all previous FL baseline methods, in terms of both PSNR and NMSE. Visual comparisons of low-count denoising before and after SA of FedFTN are shown in Figure \ref{fig:compare_sa}. We can see SA improves the image quality by further suppressing the false positive signal and improving the image resolution across different low-count levels at Institution \#1.

\begin{figure}[htb!]
\centering
\includegraphics[width=0.34\textwidth]{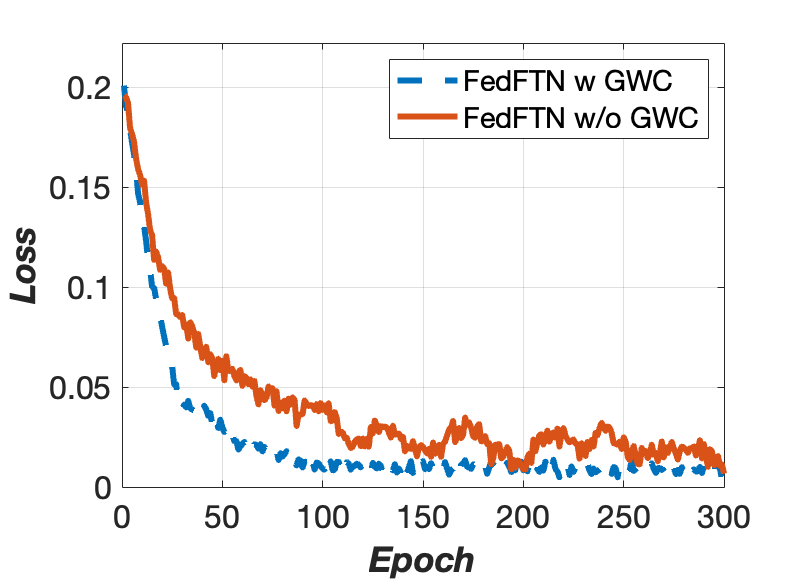}
\caption{Institution \#1's training curves of FedFTN with and without the GWC components.}
\label{fig:plot_loss}
\end{figure}

We conducted ablative studies on the Global Weight Constraint (GWC) loss that was used in our FedFTN framework to stabilize the training process. The results were summarized in Table \ref{tab:comp_gwc}, which included the image quality analysis for all low-count levels across all three institutions. Adding the GWC component consistently improved image quality compared to using FedFTN without GWC. For instance, at Institution \#1, the PSNR for the 5\% PET imaging increased from $27.11$ to $27.24$, from $28.82$ to $28.96$ for the 10\% PET imaging, and from $30.71$ to $30.82$ for the 20\% PET imaging. Figure \ref{fig:compare_gwc} provided visual comparisons from multiple institutions. We observed that adding GWC helps stabilize the FedFTN training process and helps maintain the subtle true positive signals within the kidneys, as seen in the denoised images from Institutions \#1 and \#2. Similarly, adding GWC results in better resolution recovery for clustered thin anatomic structures, such as ribs, as demonstrated in the Institution \#3 comparison. While the quantitative and qualitative improvements are incremental, as shown in Figure \ref{fig:plot_loss}, we can observe that FedFTN with GWC can provide faster and more stable training loss convergence, as compared to the one without GWC, demonstrating the GWC's benefits during FL training.

\section{Discussion} 
% Discussion points:
% 1. Discussion on our idea: FTN to enable both personalized network and unified denoising for multi-low-count levels; gwc loss to stabilize the training

In this work, we developed a novel personalized federated learning approach, called FedFTN, for multi-institutional low-count PET denoising. Specifically, we proposed to use a deep Feature Transformation Network (FTN) that is kept at the local institutions and takes the low-count level as input, to transform the intermediate feature outputs from the globally shared denoising network. There are several key advantages of this design. First of all, with different FTNs modulating the denoising network features at different local sites, we can personalize the denoising network, adapting to different low-count PET data with different distributions caused by differences in scanners, pre-/post-processing protocols, etc. Second, unlike previous personalized FL modules for image classification \citep{li2021fedbn}, MRI synthesis \citep{elmas2022federated}, and 2D SVCT reconstruction \citep{yang2022hypernetwork} that use mapper sub-network directly generate scalars for feature map channel-wise multiplication and addition, our FTN first squeezes the feature map into a latent representation and fuse with the site and subject-specific latent vector before re-excitation. In Table \ref{tab:comp_methods}, we found our FedFTN can provide better reconstruction performance than FedBN \citep{li2021fedbn} which also uses a locally kept module but generates normalization parameters for direct feature map affine transformation. Third, each institution often has multiple low-count protocols, thus requiring the denoising network to be able to adapt to inputs with different low-count levels. Instead of blindly inputting the original low-count reconstruction into the network without the knowledge of the low-count level, the FTN-modulated denoising network takes the information of the low-count level as additional input, thus enabling dose-level-aware denoising. Please also note that because individual FTN network is kept locally at each institution, the institution can define its own low-count levels. The sites do not need to share the exact same set of levels and no information about the low-count level is shared between institutions. Even though it is possible that better performance could be achieved if different sites share identical and matched low counts, having identical low counts across different sites based on different scanners, different injection protocols, and different processing/reconstruction protocols, may not be easily realizable in real-world FL scenarios. Lastly, we also proposed a Global Weight Constraint (GWC) loss that regularizes the denoising network parameters not to have strong deviation over the aggregated parameter during the training at local sites, helping stabilize the federated learning process, and thus improving the final personalized denoising performance at each site. 

% Third, unlike previous FL denoising/reconstruction methods that require specific reconstruction architecture, i.e. encoder-decoder structure \citep{guo2021multi,feng2022specificity}, for personalization at each site, our FedFTN based on FTN can be adapted to different network designs, given that we only rely on FTN to transform the features from the network to enable personalized denoising/reconstruction. 

\begin{figure}[htb!]
\centering
\includegraphics[width=0.46\textwidth]{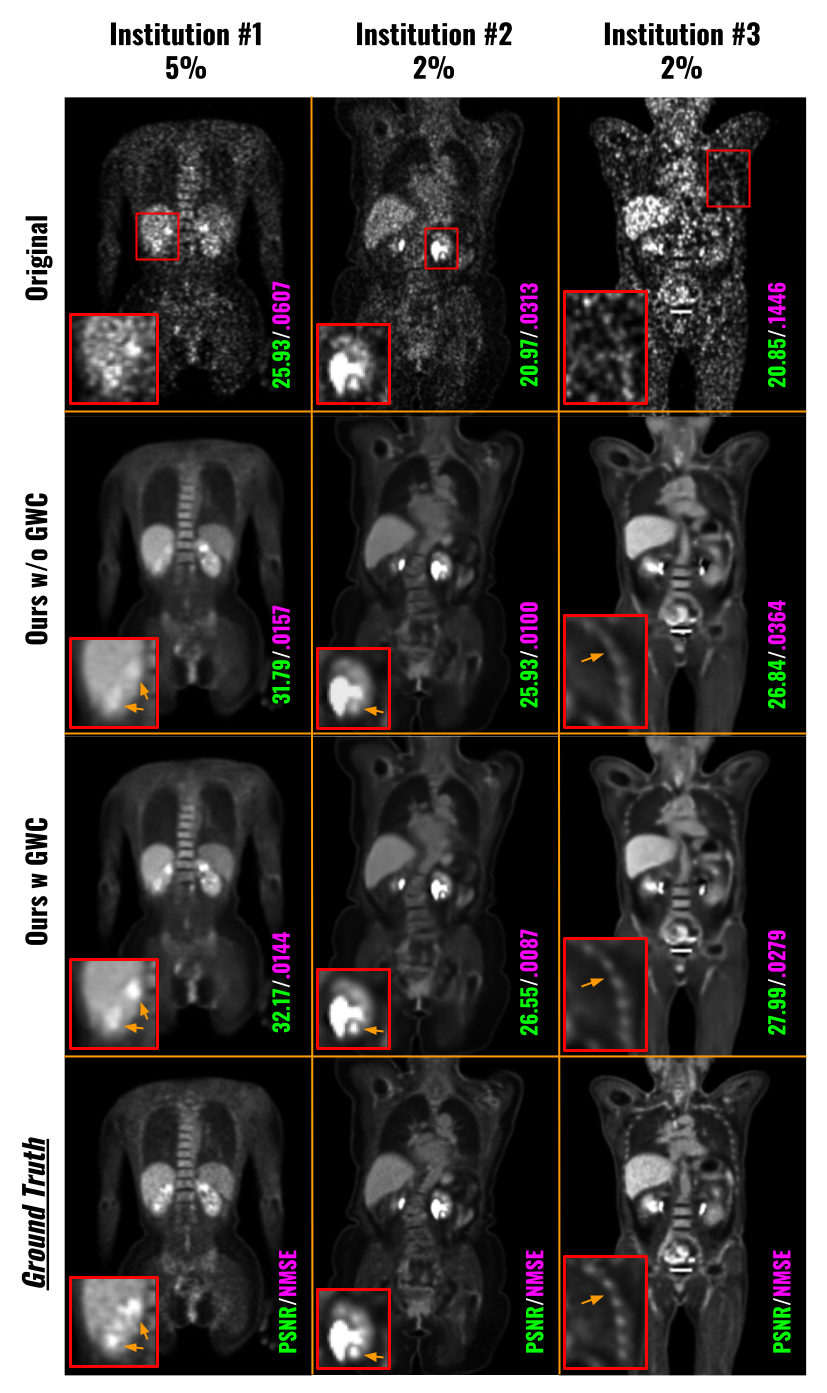}
\caption{Visual comparison of low-count denoised images from FedFTN with and without the Global Weight Constraint (GWC) loss. The PET data with the lowest dose levels at each institution are visualized here.}
\label{fig:compare_gwc}
\end{figure}

\begin{table} [htb!]
% \small
\footnotesize
\centering
\caption{Ablation study on Global Weight Constraint (GWC) loss. Quantitative comparisons of FedFTN with and without GWC loss during training are shown here.}
\label{tab:comp_gwc}
\resizebox{0.475\textwidth}{!}{
    \begin{tabular}{l|c|c|c}
        \hline
        \textbf{Institution \#1}    & 5\%    & 10\%    & 20\%   \Tstrut\Bstrut\\
        \hline
        FedFTN w/o GWC               & $27.11/.0262/.978$  & $28.82/.0178/.982$  & $30.71/.0117/.990$    \Tstrut\Bstrut\\
        FedFTN w GWC                & $27.24/.0254/.979$  & $28.96/.0172/.983$  & $30.82/.0113/.990$    \Tstrut\Bstrut\\
        \hline
        \hline
        \textbf{Institution \#2}    & 2\%    & 5\%    & 10\%   \Tstrut\Bstrut\\
        \hline
        FedFTN w/o GWC               & $25.91/.0194/.979$  & $27.59/.0137/.988$  & $28.82/.0103/.989$    \Tstrut\Bstrut\\
        FedFTN w GWC                & $26.12/.0185/.980$  & $27.80/.0130/.989$  & $29.03/.0098/.991$    \Tstrut\Bstrut\\
        \hline
        \hline
        \textbf{Institution \#3}    & 2\%    & 5\%    & 10\%   \Tstrut\Bstrut\\
        \hline
        FedFTN w/o GWC               & $26.83/.0323/.980$  & $28.87/.0223/.988$  & $30.14/.0179/.991$    \Tstrut\Bstrut\\
        FedFTN w GWC                & $26.83/.0319/.980$  & $28.91/.0219/.990$  & $30.23/.0176/.992$    \Tstrut\Bstrut\\
        \hline
    \end{tabular}
    }
\end{table}

% 2. Discussion on results: significantly outperforms previous methods, why - transformation at all feature, more flexibility on where to put in the network, gwc helps stabilize the training, local fine-tuning further improves the results

We collected large-scale real-world low-count PET data from three different institutions in the U.S.A., Europe, and China, to validate our method. From our experimental results, we demonstrated the feasibility of using our FedFTN for collaborative training without sharing data, while enabling personalized low-count PET denoising at different institutions. First, as we can observe from Table \ref{tab:comp_local} and Figure \ref{fig:compare_local}, even though training a denoising model from scratch using local data with limited diversity can generate reasonable denoising performance and potentially avoid domain shift issues, our FedFTN can provide significantly better denoising results. For example, as shown in the last row of Table \ref{tab:comp_local}, our FedFTN demonstrated superior PSNR and NMSE values as compared to the locally trained models across all the low-count levels at all the institutions. This is mainly due to the fact that the FedFTN utilizes all the institutional data with a wider spectrum of data diversity for collaborative learning while using the FTN modulation to mitigate the domain shift issues. Second, as we can see from Table \ref{tab:comp_methods} and Figure \ref{fig:compare_methods}, our method generating personalized FTN-modulated denoising networks for individual institutions can consistently outperform previous FL reconstruction methods that either only produce one global model (FedAvg) or deploy personalize FL strategies (FedSP and FedHyper). Using an identical backbone reconstruction/denoising network, our FedFTN achieved the best image quality over all the previous FL  baselines, in terms of both PSNR and NMSE with statistical significance, as reported in the second last row of Table \ref{tab:comp_methods}. Further fine-tuning the personalized denoising models from FedFTN with local data slightly boosted the performance. In addition, we found that adding the Global Weight Constraint (GWC) loss helped stabilize the FedFTN and improved the image quality at all low-count levels at all institutions, as demonstrated in Figure \ref{fig:compare_gwc} and Table \ref{tab:comp_gwc}.

% 3. Discussion on limitations: only test on FDG data / only 3 sites included in this work / no evaluation on important pathology region and downstream tasks / did not test on more advanced reconstruction architectures.

The presented work also has limitations with several potential improvements that are the subjects of our future studies.   First, our study only considered three institutions and all with \textsuperscript{18}F-FDG tracer.   While there are other PET tracers that could be used for specific applications, \textsuperscript{18}F-FDG is still the most commonly used PET tracer in clinical practice and thus is the primary focus of our study. Furthermore, our FedFTN framework can be flexibly adjusted to different numbers of institutions, and potentially adapted to multi-tracer PET scenarios. Specifically, we could incorporate the tracer type, as well as other dose and patient information, as additional inputs to the FTN, which would also allow the FTN to transform the features in the denoising network depending on the input tracer type. In fact, we believe that including more diverse tracer types with expanded training data with more diverse data representation would potentially further improve our performance. In this work, we have already shown that we can use FTN to adapt and unify those different low-count distributions. However, we believe expanding to include multi-tracer multi-institutional data is an important future direction to validate any conclusion. Second, we only evaluated the overall image quality based on image quality metrics, i.e. PSNR/NMSE/SSIM, which use full-count PET as ground truth. The difference between our FL method and locally trained baselines is in the order of 1-2 dB PSNR which implies a significant image quality improvement with our FL strategy as compared to the local training strategy, which can be observed from Table \ref{tab:comp_local} and Figure \ref{fig:compare_local}. While the image quality metrics improvements from our FedFTN as compared to the prior SOTA FL methods is in smaller magnitude, the image quality improvements are reflected on more detailed regions, as shown in Figure \ref{fig:compare_methods}. We believe this kind of improvement could potentially lead to more accurate disease quantification, e.g. lesion radiomic, cardiac function, etc. However, how will such improvement over the prior SOTA FL method be reflected in clinically relevant tasks is an important direction for our following clinical investigation. Given PET has extensive clinical applications in oncology, cardiology, and neurology, our future work also includes evaluations of how the denoised image impacts the downstream tasks, such as the impacts on staging and therapy response, and human experts' evaluations on these clinical tasks. Lastly, for a fair comparison and to demonstrate the idea, we performed all our experiments using a simple UNet as the backbone denoising network. However, our method could be adapted with more advanced restoration network structures. For instance, we could use more advanced network designs, such as cascade-based, transformer-based, and multistage-based reconstruction networks \citep{zhou2021limited,zhou2022dudodr,zhou2020dudornet,zhou2022dudoufnet,shan2019competitive}, in our FedFTN. Specifically, we can use FTN to transform and modulate the intermediate features in these networks, thus enabling personalized FL with these networks. Deploying these networks in our FedFTN could potentially further improve our performance and will also be an important direction for our future studies.

\section{Conclusion} 
Our work proposes an innovative personalized federated learning method, named FedFTN, for multi-institutional low-count PET denoising. Our method utilizes a deep feature transformation network to generate a personalized low-count PET denoising model for each institution by modulating a globally shared denoising network. During the federated learning process, the FTN remains at the local institutions and is used to transform intermediate feature outputs from the shared denoising network, thus enabling personalized FL denoising. We utilized a large-scale dataset of multi-institutional low-count PET data from three medical centers located across three continents to validate our method. Our experimental results showed that the FedFTN provides high-quality low-count PET denoised images, outperforming previous baseline FL methods across all low-count levels at all three institutions. We believe our proposed methods could potentially be adapted to other deep learning-based medical imaging challenges where collaborative training without data sharing is needed to improve the medical image quality. 

% -------------------------------------------------------------------
\section*{Acknowledgments}
This work was supported by the National Institutes of Health (NIH) grant R01EB025468 and grant R01CA275188. Parts of the data used in the preparation of this article were obtained from the University of Bern, Department of Nuclear Medicine and School of Medicine, Ruijin Hospital. As such, the investigators contributed to the design and implementation of DATA and/or provided data but did not participate in the analysis or writing of this report. A complete listing of investigators can be found at: “https://ultra-low-dose-pet.grand-challenge.org/Description/”

% -------------------------------------------------------------------
\section*{Declaration of Competing Interest}
The authors declare that they have no known competing financial interests or personal relationships that could have appeared to influence the work reported in this paper.

% -------------------------------------------------------------------
\section*{Credit authorship contribution statement }
\textbf{Bo Zhou}: Conceptualization, Methodology, Software, Visualization, Validation, Formal analysis, Writing original draft.
\textbf{Huidong Xie}: Data preparation, Methodology, Writing - review and editing.
\textbf{Qiong Liu}: Data preparation, Writing - review and editing.
\textbf{Xiongchao Chen}: Data preparation, Writing - review and editing.
\textbf{Xueqi Guo}: Data preparation, Writing - review and editing.
\textbf{Zhicheng Feng}: Results analysis, Writing - review and editing.
\textbf{Jun Hou}: Results analysis, Writing - review and editing.
\textbf{S. Kevin Zhou}: Writing - review and editing.
\textbf{Biao Li}: Data preparation, Writing - review and editing.
\textbf{Axel Rominger}: Data preparation, Writing - review and editing.
\textbf{Kuangyu Shi}: Data preparation, Writing - review and editing.
\textbf{James S. Duncan}: Writing - review and editing, Supervision.
\textbf{Chi Liu}: Conceptualization, Writing - review and editing, Supervision.

%%Harvard
% \clearpage
\bibliographystyle{model2-names.bst}\biboptions{authoryear}
\bibliography{refs}

\end{document}